\documentclass[conference]{IEEEtran}
\IEEEoverridecommandlockouts
\usepackage{cite}
\usepackage{amsmath,amssymb,amsfonts}
\usepackage{comment}
\usepackage{graphicx}
\usepackage{textcomp}
\usepackage{xcolor}
\usepackage{placeins}
\usepackage{listings}
\usepackage{url}
\usepackage[hidelinks]{hyperref}
\usepackage{graphicx}
\usepackage{subcaption}

\usepackage[utf8]{inputenc}
\usepackage[T1]{fontenc}

\begin{document}

\title{Bring Cloud-Native Networking to FreeBSD Jails:\\ Porting Calico from Linux}

\author{\IEEEauthorblockN{Soma Sakaguchi\IEEEauthorrefmark{1},
    Shintaro Suzuki\IEEEauthorrefmark{1},
    Yuki Nakata\IEEEauthorrefmark{2}{\raisebox{0pt}[0pt][0pt]{\textsuperscript{\footnotesize ,}}}\IEEEauthorrefmark{1},
    Katsuya Matsubara\IEEEauthorrefmark{1}}
  \IEEEauthorblockA{\IEEEauthorrefmark{1}Future University Hakodate, Hokkaido, Japan}
  \IEEEauthorblockA{\IEEEauthorrefmark{2}SAKURA internet Inc., Osaka, Japan}
  \IEEEauthorblockA{Email: \{g2124015, g2123032\}@fun.ac.jp, y-nakata@sakura.ad.jp, matsu@fun.ac.jp}
}

\maketitle

\begin{abstract}
While Linux containers benefit from a rich ecosystem of Container Network Interface (CNI) plugins such as Calico and Cilium, FreeBSD Jails lacked native, standardized solutions for advanced networking.
As a result, users have relied on manual, error-prone configurations, limiting scalability and integration with modern orchestration platforms.
This paper introduces porting Calico, a widely adopted CNI plugin, to the FreeBSD operating system.
Our approach replaced Linux-specific components with FreeBSD-native equivalents: iptables with ipfw, netlink sockets with routing sockets, and Linux network namespaces with FreeBSD's vnet.
This work aims to bridge the gap between Linux and BSD container networking.
By providing scalable, policy-driven capabilities that are compatible with existing Linux-centric infrastructure, FreeBSD can participate more fully in the broader container networking ecosystem.
\end{abstract}

\section{Introduction}
The container virtualization technology has been used as a lightweight application execution environment in cloud services\cite{cncf-survey}.
In cloud-native, a methodology for building and running in the cloud, applications are designed by decomposing into microservices, and containers isolate each microservice\cite{microservice-2}.
When users access a cloud-native application, frequent inter-container communications occur between microservices through network communication such as HTTP or gRPC.
Consequently, network processing dominates a significant portion of the service's overall processing\cite{frequently-communication-1}.
Thus, container networking is a crucial component of a cloud platform.

Modern container networking technology is largely based on Linux, whereas FreeBSD has not kept pace with the advancements seen in contemporary cloud-native stacks.
FreeBSD has some container networking features, such as VNET and Container Network Interface (CNI).
VNET isolates network resources between jails and the host.
However, it provides less flexible isolation than the network resource isolation features available in Linux.
CNI is the standard container networking specification for Linux, and basic CNI plugins have been partially ported to FreeBSD.
Unfortunately, it does not support common cloud configurations in which containers are deployed across multiple nodes.
We consider that to practically utilize FreeBSD as a cloud computing platform, it is essential to implement a container networking mechanism on FreeBSD that is equivalent to Linux's CNI.

We have ported Calico\cite{calico} to FreeBSD.
Calico is one of the de facto standard CNI plugins in Linux\cite{calico-standard}.
We identified technical challenges arising from differences in networking functionality between Linux and FreeBSD and devised techniques to address them. We implemented Calico modules that allow FreeBSD's networking functions to be accessed through interfaces equivalent to those on Linux.
Finally, we ensured that our Calico in FreeBSD could set up network configurations for container-to-container communication on a single node, container-to-container communication across multiple nodes, and container-to-internet communication.
These are equivalent to the common container networking configured by Linux's original Calico.

%
%

\section{Our Motivation}
\begin{figure}[t]
  \includegraphics[width=\hsize]{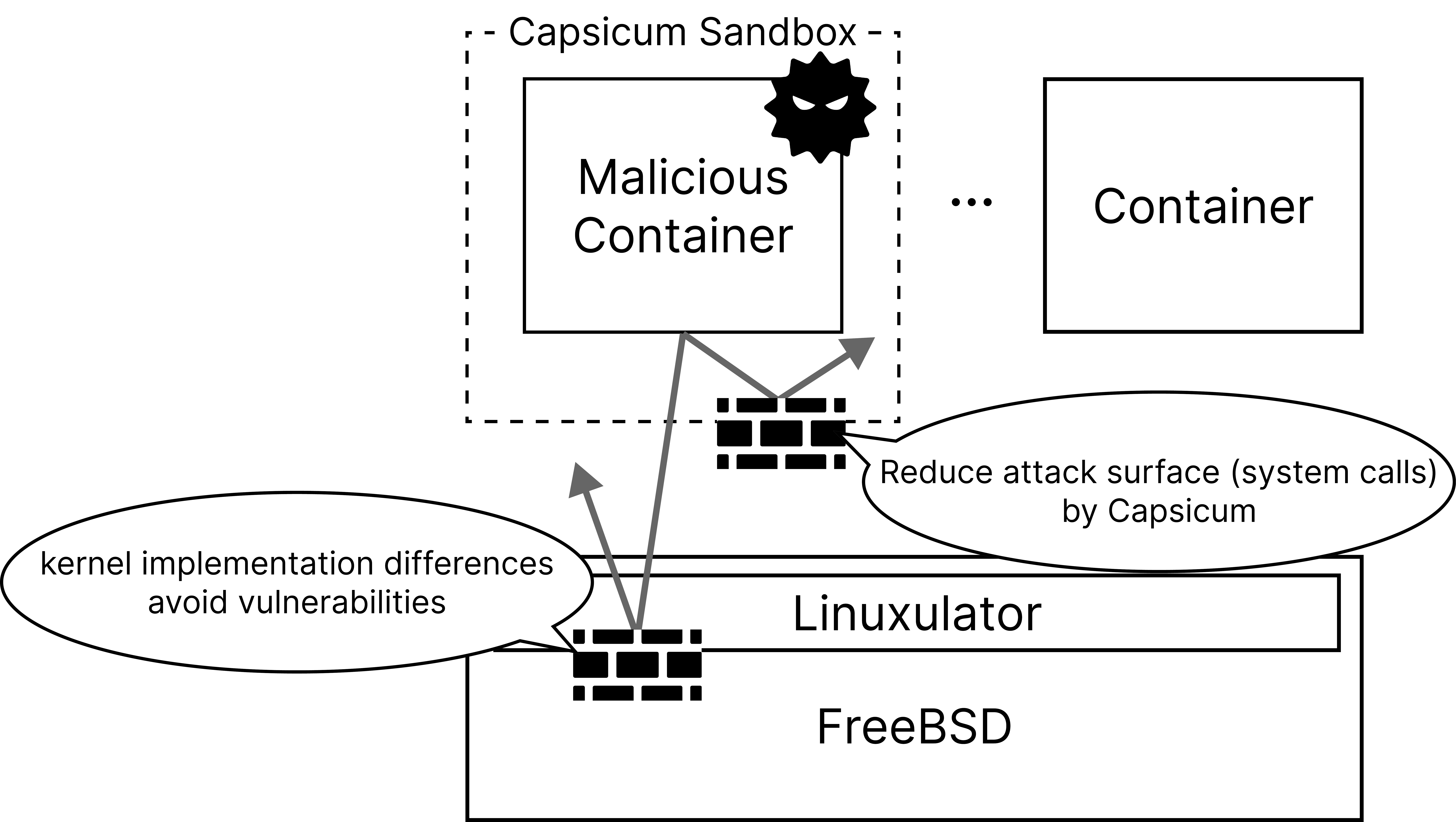}
  \caption{Overview of Secure Container on FreeBSD}
  \label{fig:runcap}
\end{figure}

The contributions of this paper are motivated by our long-term goal of making the secure container on FreeBSD practical as a cloud application execution environment.
As shown in Fig. \ref{fig:runcap}, we have proposed the secure container approach that ports a Linux container execution environment onto FreeBSD in order to avoid attacks that exploit OS kernel vulnerabilities while preserving the lightweight nature of containers as much as possible \cite{runcap}.
Compared with conventional secure container approaches, such as those that introduce additional isolation through virtual machines \cite{kata-container} or those that isolate the OS kernel using a user-space kernel \cite{gvisor}, the proposed method achieves attack mitigation with lower overhead.
While our secure container leverages FreeBSD Jail, Linuxulator and Capsicum to mitigate kernel-level attacks with low overhead, its applicability has been limited without a container networking mechanism equivalent to those used in modern cloud-native platforms.
In particular, the lack of support for multi-node container networking has been a significant obstacle to deploying microservice-based applications.

By enabling Calico on FreeBSD, this paper addresses this missing component and allows secure containers to participate in realistic cloud networking environments.
This makes it possible to deploy cloud-native applications, including microservice architectures, on top of our secure container platform.
Furthermore, because our Calico implementation replaces Linux-specific mechanisms such as Network Namespace, Netlink, and iptables with FreeBSD-native counterparts, it also provides an opportunity to study how differences in OS networking implementations can contribute to mitigating network-related attacks.
Through this work, we aim to preserve the robustness of the secure container even as networking introduces a new attack surface.

\section{Linux Container Networking}
\label{sec:linux-networking}

\begin{figure}[t]
  \includegraphics[width=\hsize]{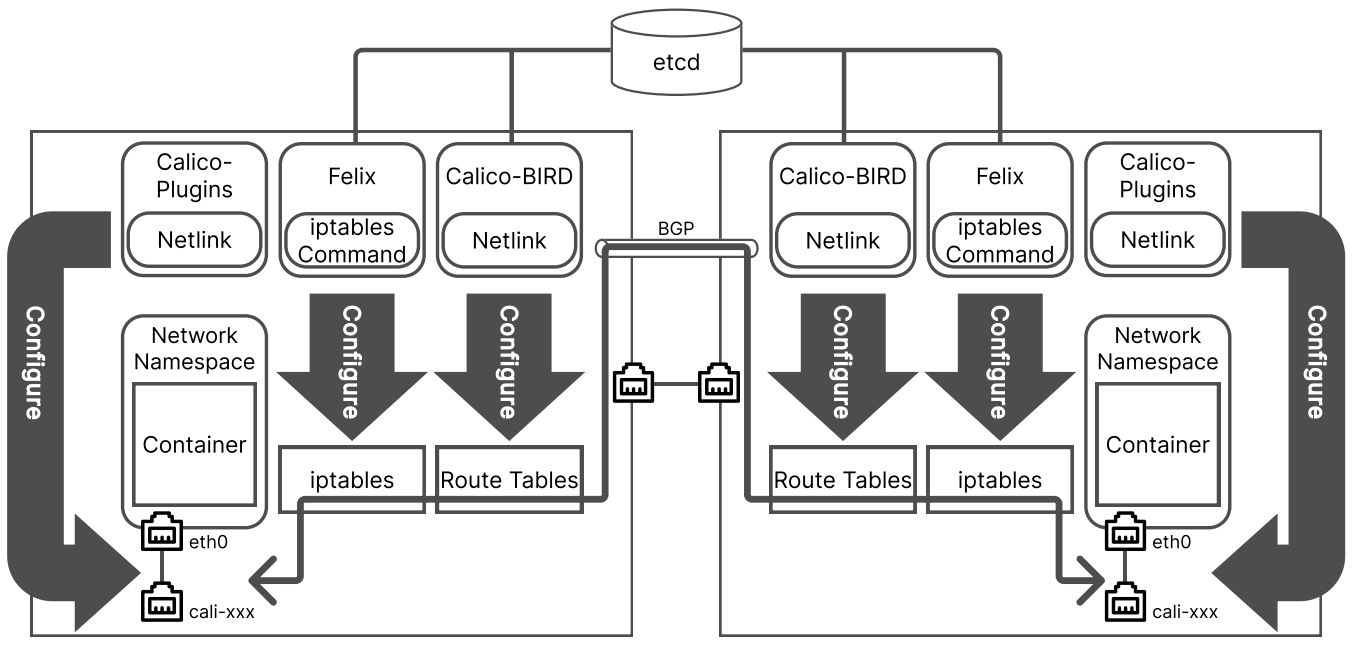}
  \caption{Calico Components for Linux Container Networking}
  \label{fig:calico}
\end{figure}

CNI provides CNI plugins to configure network resources inside containers.
CNI plugins ensure container network connectivity by using OS functions such as Network Namespaces, Netlink, and iptables.
Network Namespace isolates network resources between processes.
Each container can use its own network resources independently of the host by separating network resources from the host.
Netlink configures network resources.
A host or containers configures its link devices and IP addresses by sending requests to the kernel through Netlink.
Iptables provides packet filtering and network address translation (NAT) functionality.
It controls packets related to container communication and serves as a firewall, as well as supporting communication with external networks.

Calico is a de facto standard CNI plugin \cite{calico-standard}.
Calico enables container-to-container communication on a single node, container-to-container communication across multiple nodes, and container-to-Internet communication.
The functional components of Calico are shown in Fig. \ref{fig:calico}.
Container-to-container communication on a single node is achieved through Network Namespace management by the container runtime, and through the assignment of link devices and IP addresses and control of routing tables by the Calico plugin.
Container-to-container communication across multiple nodes is achieved by route advertisement to different nodes using Border Gateway Protocol (BGP) via Calico-BIRD.
Communication between containers and external networks is achieved by applying filtering and NAT rules using Felix.

%

\section{Differences between Linux and FreeBSD}
This section examines the current networking technologies in FreeBSD and clarifies the technical challenges involved in achieving Linux-equivalent container networking on FreeBSD.

\subsection{Differences in Network Isolation}
FreeBSD jail provides process isolation functionality equivalent to that of Linux containers, and VNET supports isolating network resources between jails.

However, VNET is not compatible with Linux Network Namespaces and does not allow a single VNET instance to be shared among multiple jails.
As a result, container deployment models such as sidecar container architectures, in which auxiliary containers are placed alongside a primary container, or configurations like Kubernetes Pods that treat multiple containers as a single minimal unit, cannot operate correctly.
To support architectures in which network resources are shared among multiple containers, it is necessary to extend the functionality of VNET.

\subsection{Differences in Network Resource Configuration}
In FreeBSD, network configuration is performed using routing sockets, the ioctl system call, and Netlink.
Routing sockets provide routing table manipulation functionality that is compatible with the Netlink interface, enabling the addition and modification of routing information.
Similar to Netlink, routing sockets use messages defined in user-space libraries to request routing table operations from the kernel.
Network resource configuration using the ioctl system call allows operations such as adding link devices and assigning IP addresses.
This is achieved by specifying the AF\_INET address family and various request codes, such as SIOCIFCREATE2 for creating new interfaces and SIOCSIFADDR for assigning IP addresses, as arguments to the ioctl system call.
Netlink has recently been ported from Linux to FreeBSD, and Linux-equivalent network configuration is partially supported.
This support covers Linux-compatible operations but excludes FreeBSD-specific network configurations such as VNET and epair.

Each of these mechanisms has technical limitations: routing sockets can only manipulate routing tables.
The ioctl system call does not provide an interface equivalent to Netlink, and FreeBSD's Netlink support is still incomplete.
Due to these limitations, achieving network resource configuration equivalent to that of Linux is difficult.
Therefore, it is necessary to enable equivalent network configuration through an interface comparable to Linux Netlink.

In addition, although several CNI plugin implementations are available for FreeBSD, they lack sufficient functionality for use in cloud computing infrastructures.
The CNI plugins available on FreeBSD are reference implementations provided by the CNI team.
Networks constructed using these plugins only support inter-container communication on a single node and do not enable inter-container communication across multiple nodes.

\subsection{Differences in Packet Filtering and NAT}
FreeBSD provides ipfw as a packet filtering and NAT mechanism.
ipfw is a framework implemented in the FreeBSD kernel that allows the application of rules for packet filtering and NAT, enabling packets to be permitted or denied based on predefined rules.
The rule structure of ipfw is simple: all rules are managed centrally and are evaluated sequentially in ascending order of their assigned rule numbers.

However, ipfw is not compatible with iptables, and differences in rule syntax and structure make it difficult to apply rules equivalent to those used in Linux.
In particular, if NAT cannot be applied, containers whose network resources are isolated from the host experience difficulty communicating with external networks and are unable to access resources on the Internet.
Therefore, to enable the application of rules equivalent to those of iptables, it is necessary to leverage the functionality of ipfw to reproduce iptables rules.

\section{Our Approach}
\label{sec:approach}
\begin{figure}[t]
  \includegraphics[width=\hsize]{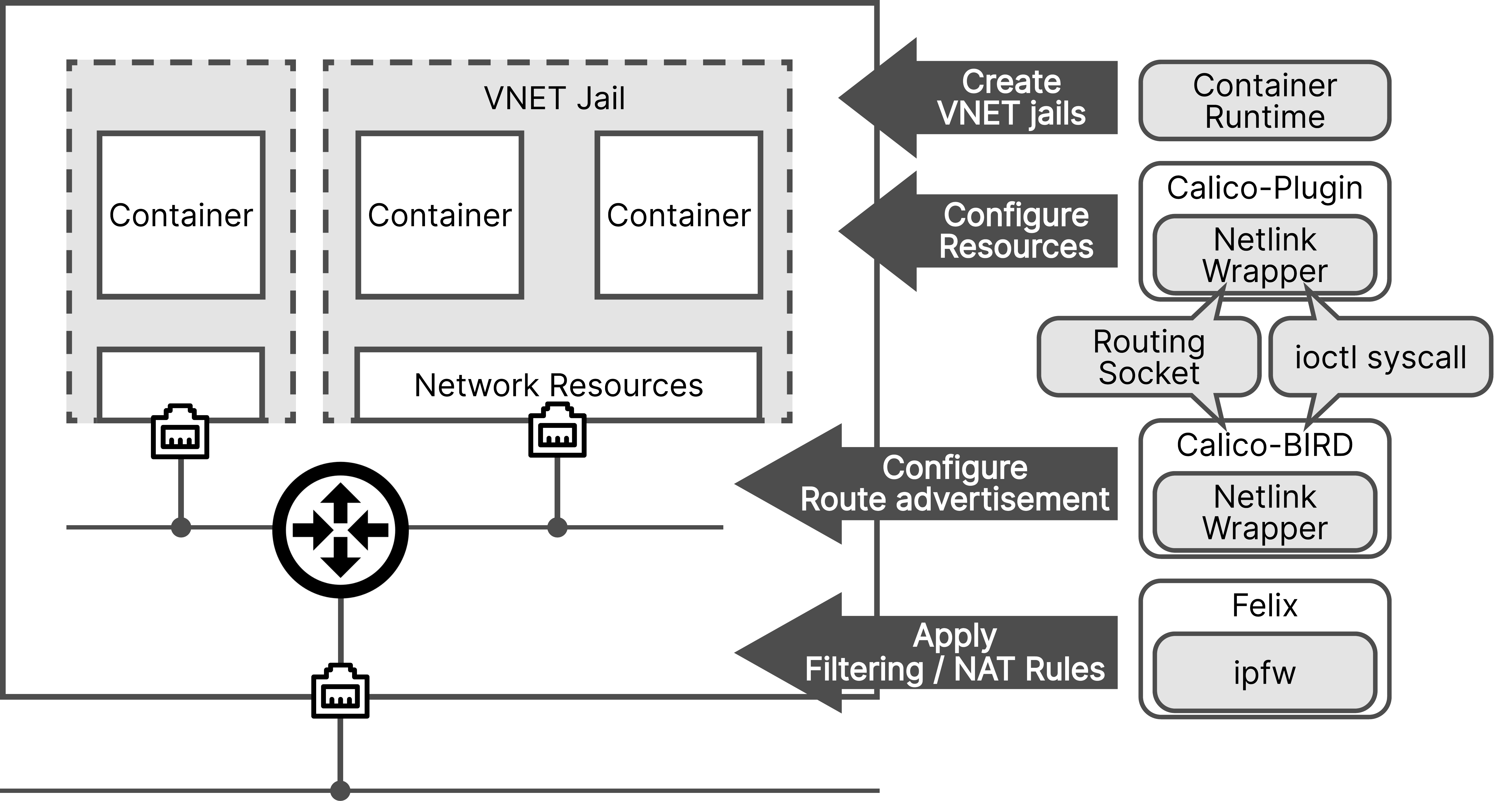}
  \caption{Architecture of Calico in FreeBSD}
  \label{fig:approach}
\end{figure}

Our approach enables Calico to operate on FreeBSD by substituting Linux-specific network configuration features with their FreeBSD counterparts.
Among the features used by Calico, several networking technologies are Linux-dependent and therefore unavailable on FreeBSD, including Linux Network Namespaces, Netlink, and iptables.
As shown in Fig. \ref{fig:approach}, this approach realizes equivalent functionality by mapping these Linux-specific mechanisms to corresponding FreeBSD facilities.
We substitute iptables with ipfw, Netlink with routing sockets and ioctl system call, and adopt FreeBSD's VNET virtualized network stack in place of Linux Network Namespaces.

To achieve functionality equivalent to Linux Network Namespaces, our approach enables network isolation on FreeBSD through a nested Jail construction mechanism in the container runtime and supporting libraries.
The container runtime achieves network isolation by creating Network Namespaces on Linux.
Although VNET provides functionality similar to Network Namespaces, it does not allow network resources to be shared among multiple containers.
To address this limitation, our approach places multiple containers inside a single VNET Jail as nested Jails, allowing them to share the same network resources.
This design enables network resource sharing semantics equivalent to those of Linux Network Namespaces.
The container runtime supports the creation and sharing of VNET Jails, and a compatibility library provides network resource operations within these Jails.
Using the jexec command, the runtime constructs nested Jails by issuing Jail-creation commands from inside an existing VNET Jail.
The compatibility library provides an interface compatible with the Linux Network Namespace library~\cite{netns-package} by replacing the UNSHARE system call with the JAIL\_SET system call.

For network resource configuration, our approach provides a wrapper library that exposes a Linux Netlink–compatible interface.
Internally, this library translates Netlink operations into equivalent FreeBSD mechanisms based on routing sockets and the ioctl system call.
FreeBSD offers several mechanisms for operating on networking resources, including routing sockets, the ioctl system call, and partial support for Netlink.
However, each mechanism has technical limitations: routing sockets operate only on routing tables, the ioctl system call does not provide a Netlink-compatible interface, and FreeBSD's Netlink support remains incomplete.
The wrapper library abstracts these differences and enables Calico components to manipulate networking resources on FreeBSD in a manner consistent with Linux behavior.
This library is designed to be compatible with the Linux Netlink library~\cite{netlink-package}.

Packet filtering and NAT functionality are realized by adapting Felix to operate on ipfw instead of iptables.
Although ipfw provides capabilities comparable to iptables, differences in rule syntax and structure require translation.
Our approach introduces a rule translation mechanism that absorbs these differences and reproduces equivalent behavior on FreeBSD.
Felix is adapted to invoke ipfw commands in place of iptables commands.
Rule management is achieved using ipfw rule numbers instead of iptables chains.
Packet marking uses ipfw tags in place of iptables marks, and IP address management uses ipfw tables as a substitute for ipset.

\section{Implementation}
\label{sec:implementation}
We modified Calico components and implemented wrapper libraries to construct container networking by substituting Linux OS features with their FreeBSD counterparts.
We implemented a nested VNET construction mechanism in the container runtime to enable isolation equivalent to Linux Network Namespaces using VNET.
In addition, we developed a VNET library to allow Calico to perform the same operations.
As a replacement for Netlink, we implemented a wrapper library using routing sockets and the ioctl system call.
As a replacement for iptables, we modified Felix to apply rules by executing ipfw commands.

\subsection{Network Resource Sharing}
\begin{figure}[t]
	\includegraphics[width=\hsize]{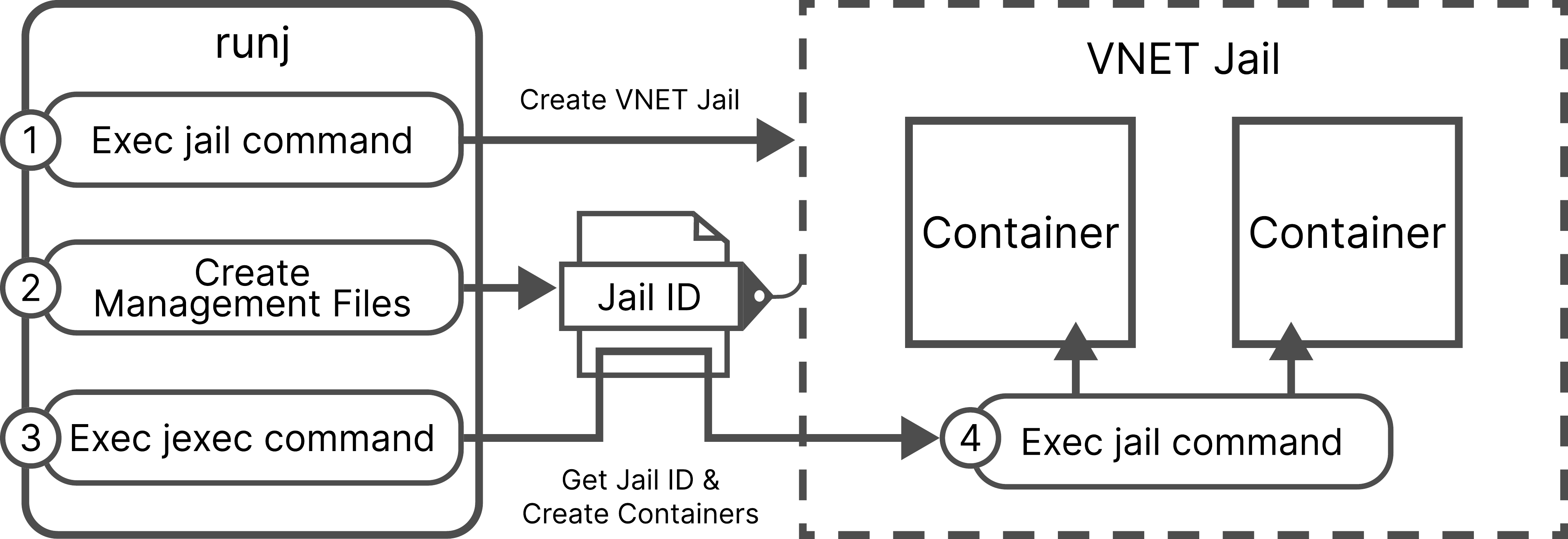}
	\caption{Nested VNET construction mechanism}
	\label{fig:vnet}
\end{figure}

To treat VNET jails as compatible with Linux Network Namespaces, we modified runj, a low-level container runtime.
An overview of the implemented mechanism is shown in Fig. \ref{fig:vnet}.
We implemented a management mechanism equivalent to that of Network Namespaces, as well as a nested jail construction mechanism.

To enable management and network isolation equivalent to Linux Network Namespaces, we implemented the creation of management files and additional configuration for VNET jails.
In Linux, when a Network Namespace is created, a management file is generated under the /proc directory or /var/run/netns, allowing the namespace to be accessed via the file system.
Similarly, in FreeBSD, we implemented a mechanism in which a management file is created under /var/run/netns when a VNET jail is created, and the corresponding jail ID is stored in this file.
In addition to the "vnet" setting, each created VNET jail is configured with children.max.
The children.max parameter defines the maximum number of child jails that can be created in a nested manner and is set to allow multiple jails to be constructed on a Network-Namespace-compatible VNET jail.

To construct nested jails, we modified runj to use the jexec command, which executes arbitrary commands inside a jail.
By specifying a jail ID and a command as arguments, jexec executes the command within the corresponding jail.
Instead of invoking the jail command directly, runj now constructs nested jails by executing the jail command inside a VNET jail using jexec with the specified VNET jail ID.

The VNET library was implemented as a compatibility library that provides an interface compatible with the widely used netns library in Linux \cite{netns-package} while replacing its internal implementation with FreeBSD VNET jails.
Through this compatibility library, each Calico component can treat VNET jails on FreeBSD in the same manner as Linux Network Namespaces while using identical function calls.
As a result, Calico can create, retrieve, and manipulate network resources within VNET jails without any modification to its implementation.

In Linux, the creation of a Network Namespace is achieved using the unshare system call.
In the VNET library, this functionality is replaced by the jail\_set system call.
Whereas the Linux netns library creates Network Namespaces using unshare, the VNET library replaces this process with the creation of a VNET jail via jail\_set.
We implemented functions in the VNET library with the same names and arguments as those defined in the netns library.
These functions invoke jail\_set so that a VNET jail is created in place of a Network Namespace.
With this design, Calico components can isolate network resources without being aware of the underlying OS differences.

In Linux, executing commands inside a Network Namespace is achieved by combining setns with fork/exec.
In contrast, the VNET library replaces this mechanism with the jail\_attach system call.
The jail\_attach system call executes the current process within the context of the specified jail, enabling arbitrary programs to run inside a VNET jail.
This allows the Calico plugin to perform network resource operations, such as creating link devices and assigning IP addresses, while operating inside a VNET jail.

\subsection{Netlink-equivalent Behavior}
\begin{figure}[t]
	\includegraphics[width=\hsize]{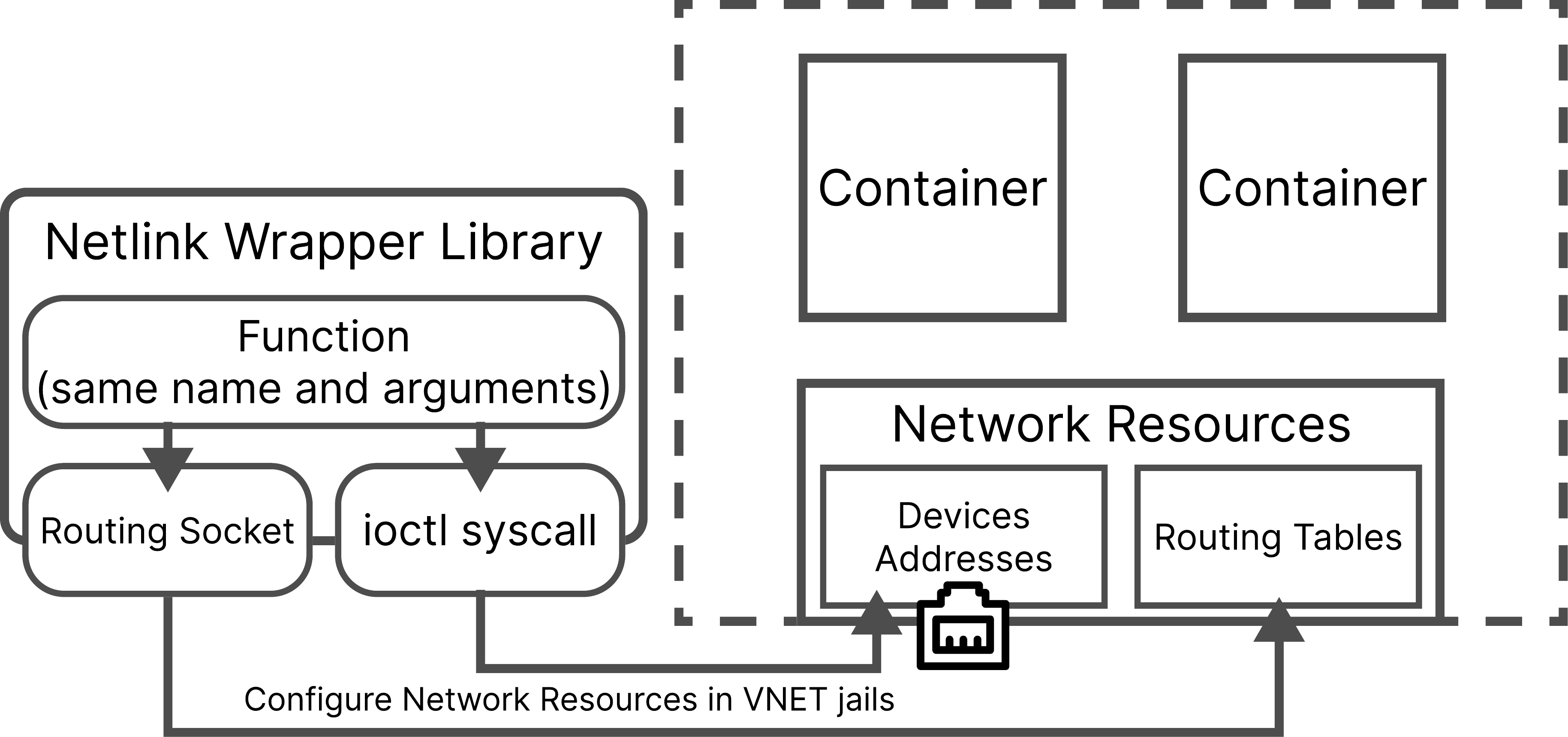}
	\caption{Netlink wrapper library}
	\label{fig:netlink}
\end{figure}

The wrapper library based on routing sockets and the ioctl system call was implemented as a compatibility library for the Netlink library used in Linux \cite{netlink-package}.
Fig. \ref{fig:netlink} illustrates the implemented mechanism.
This enables Calico components to use routing sockets and the ioctl system call in place of Netlink.

For routing control using routing sockets, we adapted the constants and data structures used by Netlink to their corresponding counterparts in routing sockets.
Although FreeBSD routing sockets, like Linux Netlink, provide a message-based interface, they differ in address families and message formats.
While Linux specifies AF\_NETLINK when creating a socket for Netlink communication, FreeBSD creates routing sockets by specifying PF\_ROUTE.
Therefore, we modified the socket creation process in the Netlink-compatible library to use PF\_ROUTE instead of AF\_NETLINK.
In addition, Netlink-specific structures such as nlmsghdr and attribute structures cannot be used directly with FreeBSD routing sockets.
Thus, we redefined these structures to match the message format of routing sockets.
As a result, Netlink messages generated by Calico can be translated into a format interpretable by FreeBSD routing sockets and transmitted accordingly.
Operations such as route addition and deletion are realized by sending requests constructed from these header structures and messages to the kernel via routing sockets.

To support link device creation and address configuration via the ioctl system call, we implemented functions with the same names and arguments as their Netlink counterparts, internally invoking ioctl.
By using requests such as SIOCIFCREATE2 and SIOCAIFADDR, the behavior of the corresponding Netlink functions was reproduced.
For virtual interface creation, we replaced veth with epair, enabling an equivalent link configuration.
Furthermore, by using the SIOCSIFVNET request to attach interfaces to VNET jails, interfaces can also be added to isolated Network Namespaces.

\subsection{iptables-equivalent Behavior}
\begin{figure}[t]
	\includegraphics[width=\hsize]{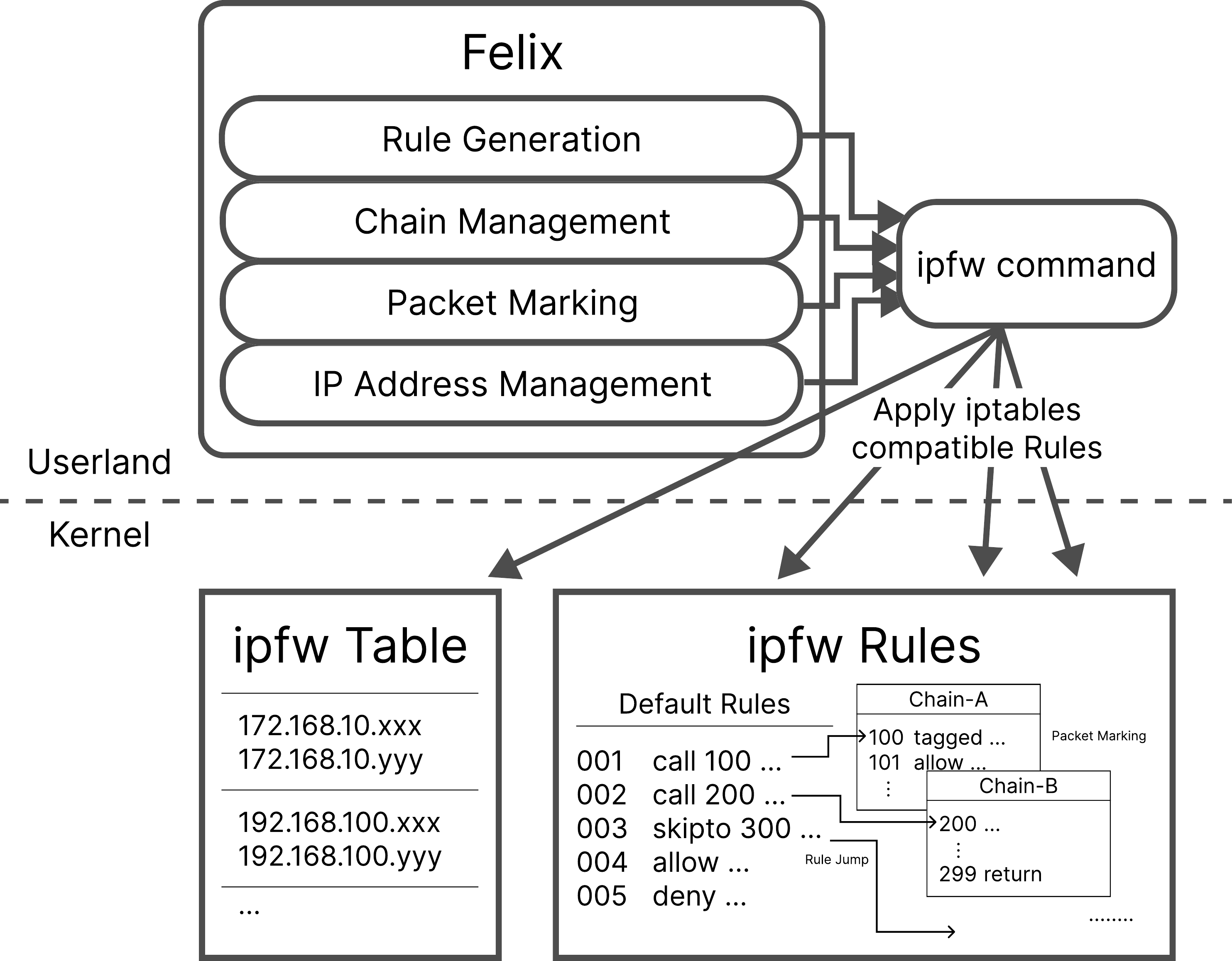}
	\caption{Applying IPFW rules from Felix}
	\label{fig:ipfw}
\end{figure}

We analyzed the syntactic differences between iptables and ipfw and modified Felix to enable the use of ipfw.
Iptables provides four main functions: packet control through rules, rule organization using Chains, packet marking using Mark, and management of IP address sets using IPSet.
While ipfw provides functionally equivalent capabilities, there are differences in terminology and mechanisms, including different rule syntax, rule management based on rule numbers, packet marking using Tag, and IP address set management using Table.

To absorb these differences, we modified Felix's execution of iptables commands so that they are replaced with equivalent ipfw functionality.
An overview of the implemented mechanism is shown in Fig. \ref{fig:ipfw}.
We implemented a mechanism that translates rule application performed via iptables commands into equivalent rule application using ipfw commands.
By converting command names, argument order, and action names to their ipfw counterparts, we were able to bridge the differences between iptables and ipfw command execution.
In addition, we reproduced the default rules applied by iptables using ipfw and modified Felix to apply these default rules at startup.

As a replacement for chains, we designed a scheme in which ipfw rule numbers are divided into fixed ranges, with each range treated as a group of rules corresponding to an iptables chain.
Rule numbers are allocated in increments of 100, and each 100-number range is managed as a single chain.
To reproduce chain jumps and returns, we leveraged the call, goto, and return actions provided by ipfw.
This design enables the reproduction of a chain-equivalent rule structure using ipfw.

As a replacement for the Mark function, we adopted the Tag mechanism.
Although Mark in iptables and Tag in ipfw differ in the bit width of values that can be assigned, the values specified in Mark by the default iptables rules generated by Calico are limited.
In the scope of Calico usage targeted in this paper, no cases requiring the full 32-bit width of Mark were observed.
Therefore, we determined that a 16-bit Tag is sufficient to provide equivalent functionality.

As a replacement for IPSet, we adopted the Table mechanism.
Both IPSet and Table allow the definition of named sets of IP addresses and the use of these sets to specify packet selection conditions.
Since there are no significant functional differences between them, IPSet operations and lookups generated by Felix can be replaced by corresponding ipfw Table operations.

\section{Operational Demonstration}
\label{fig:validate}
\begin{figure*}[t]
    \centering
    \begin{subfigure}[t]{0.32\linewidth}
        \centering
        \includegraphics[width=\linewidth]{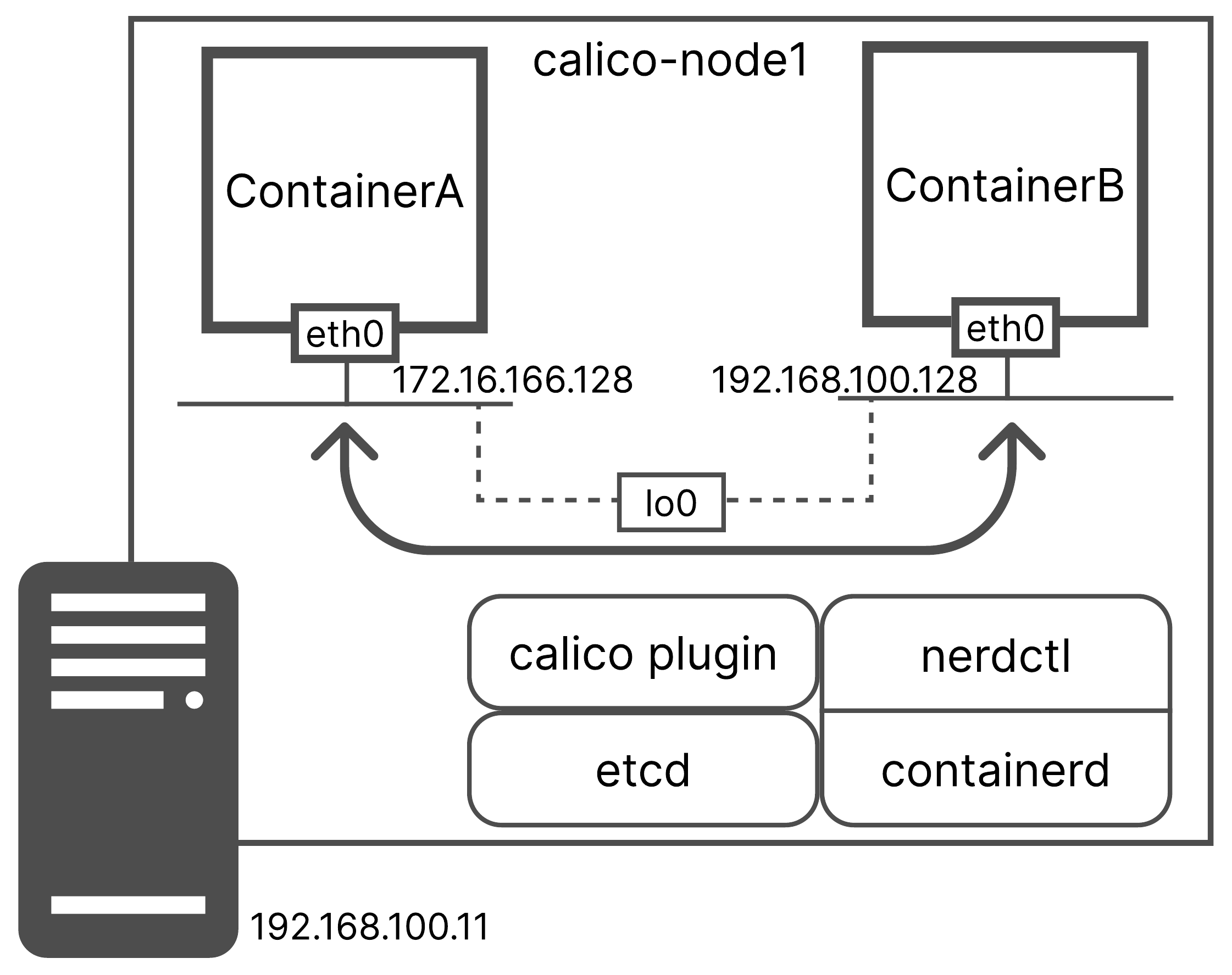}
        \caption{Container-to-container on a single node}
        \label{fig:single-node}
    \end{subfigure}
    \hfill
    \begin{subfigure}[t]{0.32\linewidth}
        \centering
        \includegraphics[width=\linewidth]{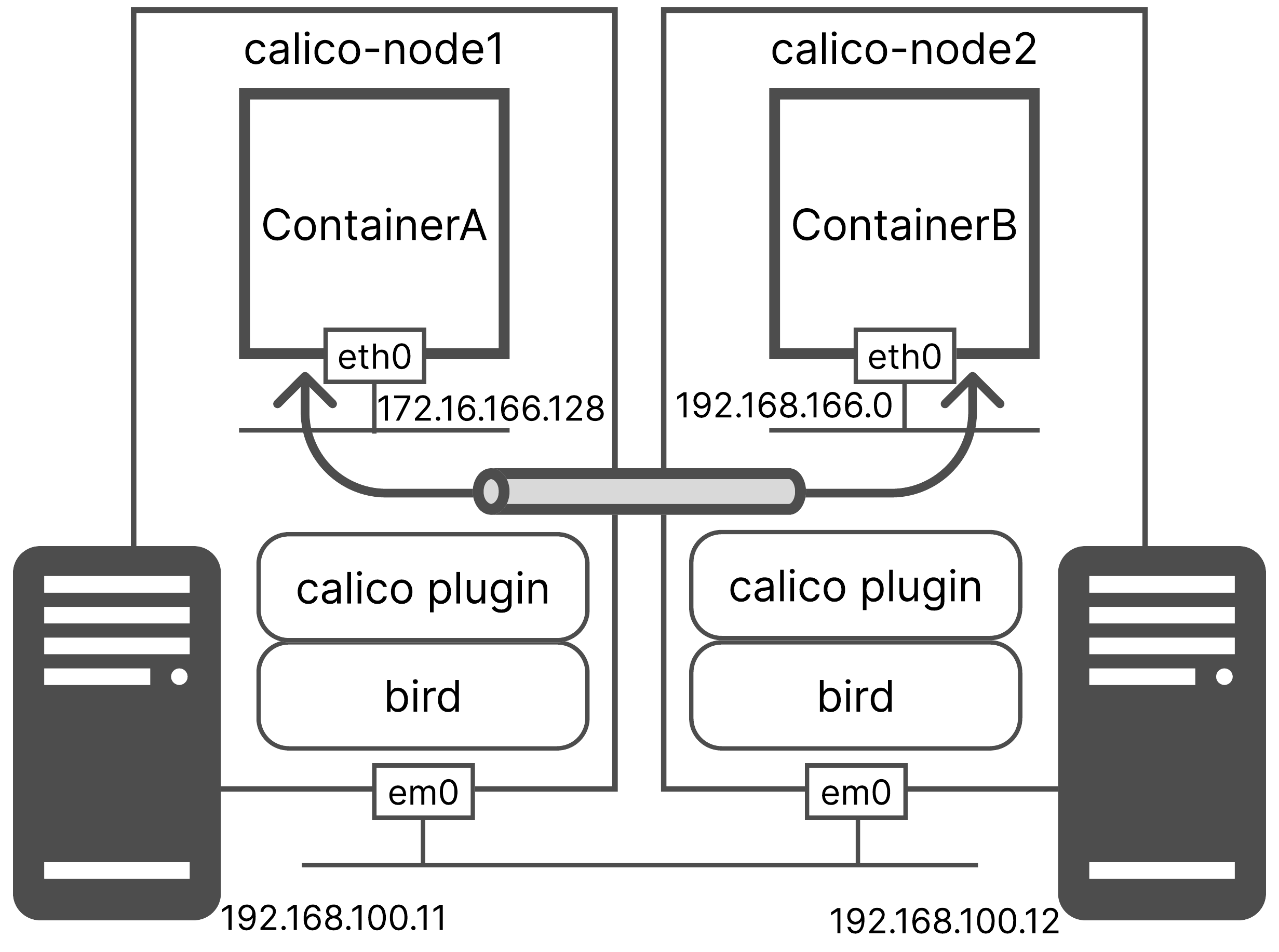}
        \caption{Container-to-container among nodes}
        \label{fig:among-nodes}
    \end{subfigure}
    \hfill
    \begin{subfigure}[t]{0.32\linewidth}
        \centering
        \includegraphics[width=\linewidth]{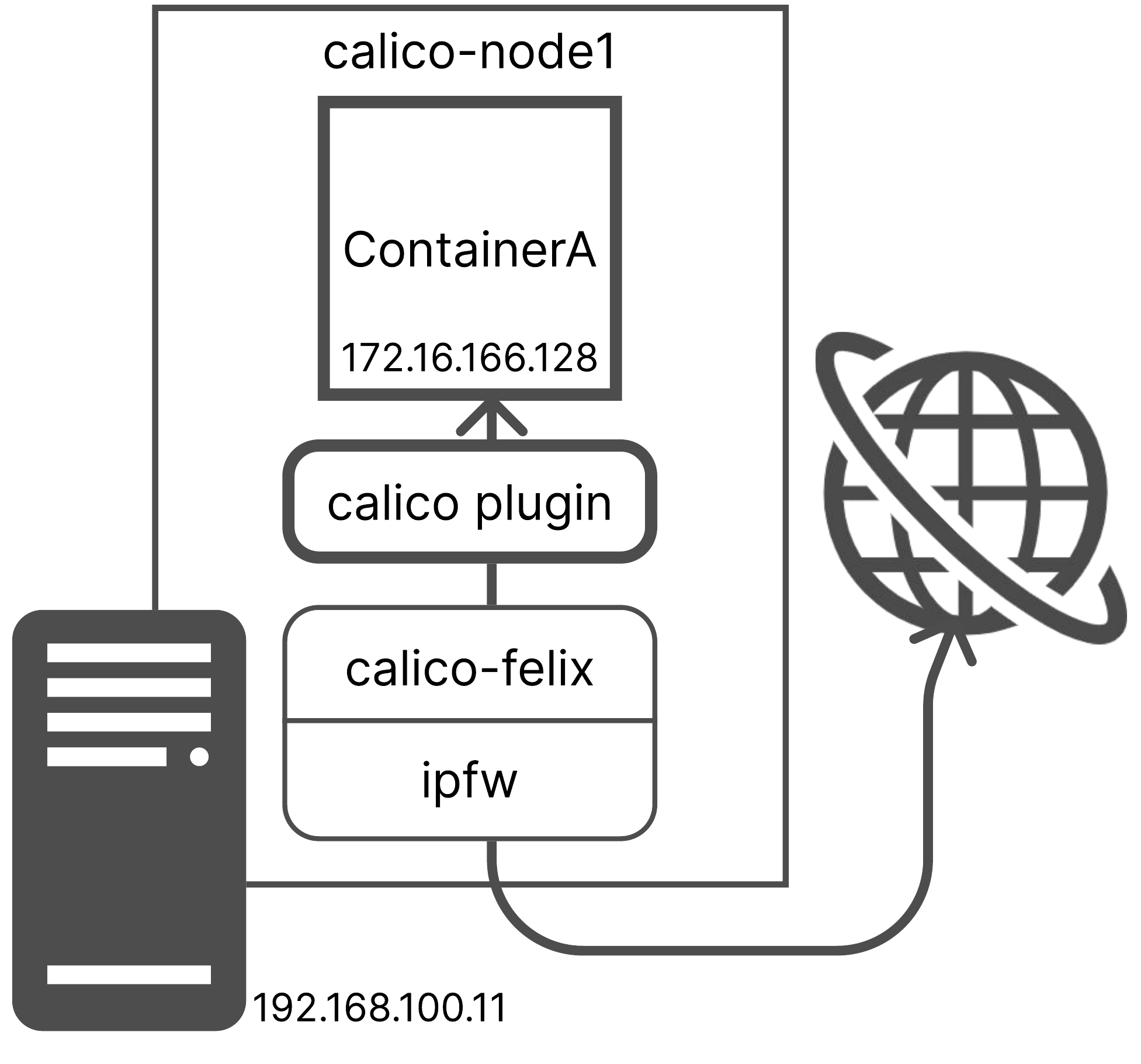}
        \caption{Container-to-internet}
        \label{fig:to-internet}
    \end{subfigure}
    \caption{Container communication supported by Calico}
	\label{fig:demo}
\end{figure*}

\begin{figure}[t]
\begin{lstlisting}[frame=single, framerule=0.5pt, basicstyle=\ttfamily\footnotesize]
apiVersion: projectcalico.org/v3
kind: IPPool
metadata:
  name: default-ipv4-ippool
spec:
  cidr: 172.16.166.0/24
  ipipMode: Never
  vxlanMode: Never
  natOutgoing: true
  disabled: false
\end{lstlisting}
\caption{container-network.yaml}
\label{fig:AB-network}
\end{figure}

\begin{figure}[t]
\begin{lstlisting}[frame=single, framerule=0.5pt, basicstyle=\ttfamily\footnotesize]
{
  "kind": "ClusterInformation",	
  "apiVersion": "projectcalico.org/v3",
  "metadata": {
    "name": "default"
  },
  "spec": {
    "datastoreReady": true
  }
}
\end{lstlisting}
\caption{Cluster Information into etcd}
\label{fig:cluster-info}
\end{figure}

\begin{figure}[t]
\begin{lstlisting}[frame=single, framerule=0.5pt, basicstyle=\ttfamily\footnotesize]
{
  "kind": "Node",
  "apiVersion": "projectcalico.org/v3",
  "metadata": {
    "name": "node1"
  },
  "spec": {
    "bgp": {
      "ipv4Address": "192.168.100.11/24"
    }
  },
  "status": {}
}
\end{lstlisting}
\caption{Node Information into etcd}
\label{fig:node-info}
\end{figure}

\begin{figure}[t]
\begin{lstlisting}[frame=single, framerule=0.5pt, basicstyle=\ttfamily\footnotesize]
router id 192.168.100.11;

protocol kernel {
  scan time 20;
  export all;
}

protocol direct {
  interface "*";
}

protocol bgp node2 {
  local as 64512;
  neighbor 192.168.100.12 as 64512;
  export all;
  import all;
}
\end{lstlisting}
\caption{BIRD configuration}
\label{fig:bird-conf}
\end{figure}

\begin{figure}[t]
\begin{lstlisting}[frame=single, framerule=0.5pt, basicstyle=\ttfamily\footnotesize]
{
  "cniVersion": "1.0.0",
  "name": "99-calico",
  "plugins": [
    {
      "datastore_type": "etcdv3",
      "etcd_endpoints": "http://127.0.0.1:2379,
      "type": "calico",
      "ipam": { "type": "calico-ipam" }
    }
  ]
}
\end{lstlisting}
\caption{Calico Plugin Configuration for Nerdctl}
\label{fig:99-calico}
\end{figure}

We describe how Calico is used in FreeBSD, assuming actual operational deployment.
We explain a typical workflow, from container startup to network configuration and communication verification.
It becomes possible to construct container networks as shown in Fig. \ref{fig:demo}.

Operators deploy Calico components and define the range of IP addresses assigned to containers using calicoctl on each node.
\begin{lstlisting}[frame=single, framerule=0.5pt, basicstyle=\ttfamily\footnotesize]
 $ sudo calicoctl apply -f ./container-network.yaml
\end{lstlisting}
They describe these settings in YAML format and store them in etcd (Fig. \ref{fig:AB-network}).
Operators can configure arbitrary address ranges according to their requirements by specifying the IP address range in the cidr field.
By registering cluster information and node information, Calico obtains an overview of the entire network configuration (Fig. \ref{fig:cluster-info}, Fig.\ref{fig:node-info}).

Operators configure BGP routing on each node using Calico-BIRD to enable container-to-container communication among multiple nodes.
They define neighboring nodes on each node and start BIRD, which exchanges container route information between nodes (Fig. \ref{fig:bird-conf}).
\begin{lstlisting}[frame=single, framerule=0.5pt, basicstyle=\ttfamily\footnotesize]
 $ sudo bird -c ./bird.conf
\end{lstlisting}
Adding protocol bgp entries enables route exchange with multiple nodes in the BIRD configuration.
As a result, containers running on different nodes can communicate directly with each other.

Operators start Felix, which applies filtering and NAT rules using ipfw to enable container-to-internet.
\begin{lstlisting}[frame=single, framerule=0.5pt, basicstyle=\ttfamily\footnotesize]
 $ DATASTORE_TYPE=etcdv3 \
   ETCD_ENDPOINTS=http://127.0.0.1:2379 \ 
   sudo -E calico-felix
\end{lstlisting}
When Felix starts, it automatically configures the required ipfw rules, allowing containers to communicate with external networks.

Operators launch containers using nerdctl and use the ported Calico as the CNI plugin.
\begin{lstlisting}[frame=single, framerule=0.5pt, basicstyle=\ttfamily\footnotesize]
 $ CNI_PATH=/opt/cni/bin \
   sudo -E nerdctl run -it --network=99-calico \
   freebsd/freebsd-runtime:14.snap /bin/sh
\end{lstlisting}
Based on the network configuration specified at container startup, the system automatically configures virtual network interfaces and routing rules (Fig. \ref{fig:99-calico}).

After launching the containers, operators verify connectivity using the ping command, confirming that container-to-container communication is possible.
In addition, they confirm that containers can also communicate with the Internet.
These operational procedures are almost identical to those used for Calico in Linux environments without Kubernetes, and existing operational know-how for Calico can be applied directly to FreeBSD environments.

\section{Conclusion}
We proposed a port of Calico to FreeBSD in order to enable FreeBSD to be used as a cloud computing platform comparable to Linux.
Standard Linux container networking is realized using Linux-specific features such as Network Namespaces, Netlink, and iptables.
To achieve equivalent container networking in a FreeBSD environment, we implemented a nested VNET jail construction method, developed a wrapper library using routing sockets and ioctl system call, and modified Felix to operate with ipfw.
Through these implementations, the port of Calico to FreeBSD was completed, and we confirmed successful communication in container-to-container on a single node, container-to-container across multiple nodes, and container-to-Internet scenarios.


\FloatBarrier
\vspace{12pt}
\bibliography{ref}

@misc{cncf-survey,
	author       = {Cloud Native Computing Foundation},
	title        = {Cloud Native Survey 2020},
	year         = {},
	month        = {},
	date         = {},
	howpublished = {\url{https://www.cncf.io/wp-content/uploads/2020/11/CNCF_Survey_Report_2020.pdf}},
	note         = {(Accessed: 2025-12-15)}
}

@article{microservice-2,
	title        = {About microservices, containers and their underestimated impact on network performance},
	author       = {Kratzke, Nane},
	journal      = {arXiv preprint arXiv:1710.04049},
	year         = {2017},
	howpublished = {\url{https://arxiv.org/abs/1710.04049}}
}

@INPROCEEDINGS{frequently-communication-1,
	author    = {Amaral, Marcelo and Polo, Jordà and Carrera, David and Mohomed, Iqbal and Unuvar, Merve and Steinder, Malgorzata},
	booktitle = {2015 IEEE 14th International Symposium on Network Computing and Applications},
	title     = {Performance Evaluation of Microservices Architectures Using Containers},
	year      = {2015},
	volume    = {},
	number    = {},
	pages     = {27-34},
	doi       = {10.1109/NCA.2015.49}
}

@INPROCEEDINGS{runcap,
	author    = {Nakata, Yuki and Suzuki, Shintaro and Matsubara, Katsuya},
	title     = {Reducing Attack Surface with Container Transplantation for Lightweight Sandboxing},
	year      = {2023},
	isbn      = {9798400703058},
	publisher = {Association for Computing Machinery},
	address   = {New York, NY, USA},
	url       = {https://doi.org/10.1145/3609510.3609820},
	doi       = {10.1145/3609510.3609820},
	booktitle = {Proceedings of the 14th ACM SIGOPS Asia-Pacific Workshop on Systems},
	pages     = {58–64},
	numpages  = {7},
	location  = {Seoul, Republic of Korea},
	series    = {APSys '23}
}

@INPROCEEDINGS{kata-container,
	author    = {Randazzo, Alessandro and Tinnirello, Ilenia},
	booktitle = {2019 Sixth International Conference on Internet of Things: Systems, Management and Security (IOTSMS)},
	title     = {Kata Containers: An Emerging Architecture for Enabling MEC Services in Fast and Secure Way},
	year      = {2019},
	volume    = {},
	number    = {},
	pages     = {209-214},
	doi       = {10.1109/IOTSMS48152.2019.8939164},
}

@misc{gvisor,
	author       = {The gVisor Authors},
	title        = {The Container Security Platform - gVisor},
	howpublished = {\url{https://gvisor.dev/}},
	month        = {},
	year         = {},
	note         = {(Accessed: 2025-11-2)}
}

@misc{calico,
	author       = {TIGERA},
	title        = {Project Calico},
	howpublished = {\url{https://www.tigera.io/project-calico/}},
	month        = {},
	year         = {},
	note         = {(Accessed: 2025-12-16)}
}

@misc{calico-standard,
	author       = {DATADOG},
	title        = {11 Facts About Real-World Container Use},
	howpublished = {\url{https://www.datadoghq.com/ja/container-report-2020/}},
	month        = {11},
	year         = {2020},
	note         = {(Accessed: 2025-11-12)}
}

@rfc{netlink,
	author    = {Salim, J. and Khosravi, H. and Kleen, A. and Kuznetsov, A.},
	title     = {RFC3549: Linux Netlink as an IP Services Protocol},
	year      = {2003},
	publisher = {RFC Editor},
	address   = {USA}
}

@misc{netns-package,
	author       = {vishvananda},
	title        = {netns},
	howpublished = {\url{https://github.com/vishvananda/netns}},
	month        = {},
	year         = {},
	note         = {(Accessed on 2025/11/15)}
}

@misc{netlink-package,
	author       = {vishvananda},
	title        = {netlink},
	howpublished = {\url{https://github.com/vishvananda/netlink}},
	month        = {},
	year         = {},
	note         = {(Accessed on 2025/11/15)}
}
\bibliographystyle{IEEEtran}
\end{document}